\documentclass[aps,reprint,prb,amsmath,amssymb,citeautoscript,
  numerical,twocolumn,showpacs]{revtex4-1}

\usepackage{graphicx} 
\usepackage{amssymb}
\usepackage{amsmath}
\usepackage{bm}
\setcitestyle{super}

\newcommand{\p}{\partial}

\newcommand{\vF}{v_{_{F}}}
\newcommand{\kB}{k_{_{B}}}
\newcommand{\epst}{\epsilon_{t}}
\newcommand{\epsb}{\epsilon_{b}}
\newcommand{\epss}{\epsilon_{s}}
\newcommand{\Ht}{H_{t}}
\newcommand{\Hb}{H_{b}}

\begin{document}

\title{Damping of Terahertz Plasmons in Graphene \\ 
  Coupled with Surface Plasmons in Heavily-Doped Substrate}

\author{A.~Satou$^{1,2}$}\email{a-satou@riec.tohoku.ac.jp}
\author{Y.~Koseki$^{1}$}
\author{V.~Ryzhii$^{1,2}$}
\author{V.~Vyurkov$^{3}$}
\author{T.~Otsuji$^{1,2}$}

\affiliation{
  $^{1}$Research Institute of Electrical Communication, 
  Tohoku University, Sendai 980-8577, Japan
  \\
  $^{2}$ CREST, Japan Science and Technology Agency,
  Tokyo 107-0075, Japan
  \\
  $^{3}$Institute of Physics and Technology,
  Russian Academy of Sciences, Moscow 117218, Russia
}

\begin{abstract}
Coupling of plasmons in graphene at terahertz (THz) frequencies
with surface plasmons in a heavily-doped substrate is studied
theoretically. We reveal that a huge scattering rate 
may completely damp out the plasmons,
so that proper choices of material and geometrical parameters are 
essential to suppress
the coupling effect and to obtain the minimum damping rate 
in graphene.
Even with the doping concentration $10^{19}-10^{20}$ cm$^{-3}$ and
the thickness of the dielectric layer between graphene and
the substrate $100$ nm, which are typical values in real graphene 
samples with a heavily-doped substrate, the increase in the damping 
rate is not negligible in comparison with the acoustic-phonon-limited
damping rate.
Dependence of the damping rate on wavenumber, 
thicknesses of graphene-to-substrate and gate-to-graphene separation,
substrate doping concentration, and dielectric constants of 
surrounding materials are investigated.
It is shown that the damping rate can be much 
reduced by the gate screening, which suppresses the field spread
of the graphene plasmons into the substrate.
\end{abstract}

\maketitle
\newpage

\section{Introduction}

Plasmons in two-dimensional electron gases (2DEGs) can be utilized for
terahertz (THz) devices.
THz sources and detectors based on compound semiconductor 
heterostructures have been extensively investigated both experimentally
and theoretically~\cite{Dyakonov1993,Dyakonov1996,Shur2000,Shur2003,
Teppe2005,Otsuji2008,Ryzhii2008c,Popov2011a}.
The two-dimensionality, which gives rise to the wavenumber-dependent
frequency dispersion, and the high electron concentration 
on the order of $10^{12}$ cm$^{-2}$ allow us to have 
their frequency in the THz range with 
submicron channel length.
Most recently, a very high detector responsivity of
the so-called asymmetric double-grating-gate structure
based on an InP-based high-electron-mobility transistor
was demonstrated~\cite{Watanabe2012}.
However, resonant detection as well as single-frequency
coherent emission have not been accomplished so far at room 
temperature, mainly owning to the damping rate more than
$10^{12}$ s$^{-1}$ in compound semiconductors.

Plasmons in graphene
have potential to surpass those in
the heterostructures with 2DEGs based on the standard
semiconductors, due to its exceptional electronic
properties~\cite{Ryzhii-JJAP-2006}.
Massive experimental and theoretical works have been done very 
recently on graphene plasmons in 
the THz and infrared regions
(see review papers 
Refs.~\onlinecite{Otsuji-JPD-2012,Grigorenko2012} 
and references therein).
One of the most important advantages of plasmons 
in graphene over those in heterostructure 2DEGs is much 
weaker damping
rate close to $10^{11}$ s$^{-1}$ at room temperature in 
disorder-free graphene suffered only from acoustic-phonon 
scattering~\cite{Hwang2008b}.
That is very promising for the realization 
of the resonant THz detection~\cite{Ryzhii2012a} and 
also of plasma instabilities,
which can be utilized for the emission.
In addition, interband population inversion in 
the THz range was predicted~\cite{Ryzhii2007b, Ryzhii2007d},
and it has been investigated 
for the utilization not only in THz lasers in the usual sense
but also in THz active plasmonic 
devices~\cite{Dubinov-JPCM-2011, Popov2012}
and metamaterials~\cite{Takatsuka2012}.

Many experimental demonstrations of graphene-based devices have
been performed on graphene samples with heavily-doped
substrates, in order to tune the carrier concentration in graphene
by the substrate as a back gate.
Typically, either peeling or CVD graphene transferred onto
a heavily-doped p$^{+}$-Si substrate,
with a SiO$_{2}$ dielectric layer in between, is
used (some experiments on graphene plasmons have adapted undoped 
Si/SiO$_{2}$ substrates~\cite{Ju-NN-2011,Strait2013}).
Graphene-on-silicon, which is epitaxial graphene 
on doped Si substrates~\cite{Suemitsu-JSSN-2009}, is also used.
For realization of THz plasmonic devices, properties of plasmons
in such structures must be fully understood.
Although the coupling of graphene plasmons to surface plasmons in
perfectly conducting metallic substrates with/without dielectric
layers in between have been theoretically
studied~\cite{Horing2009a, Yan2011}, 
the influence of the carrier scattering in a heavily-doped 
semiconductor substrate (with finite complex conductivity)
has not been taken into account so far. 
Since the scattering rate in the substrate increases as the
doping concentration increases, it is anticipated that 
the coupling of graphene plasmons to surface plasmons
in the heavily-doped substrate causes undisired increase in 
the damping rate.

The purpose of this paper is to study theoretically 
the coupling between
graphene plasmons and substrate surface plasmons
in a structure with a heavily-doped substrate 
and with/without a metallic top gate.
The paper is organized as follows. In the Sec.~II,
we derive a dispersion equation of the coupled modes of 
graphene plasmons and substrate surface plasmons.
In Sec.~III, we study coupling effect in the ungated structure,
especially the increase in the plasmon damping rate due to the
coupling and its dependences on the doping concentration,
the thickness of graphene-to-substrate separation, and
the plasmon wavenumber.
In Sec.~IV, we show that the coupling in the gated structures
can be less effective due to the gate screening. We also compare
the effect in structures having different dielectric layers
between the top gate, graphene layer, and substrate, and reveal
the impact of values of their dielectric constants.
In Sec.~V, we discuss and summarize the main results of this paper.

\section{Equations of The Model}

\begin{figure}[t]
  \begin{center}
    \includegraphics[width=8.5cm]{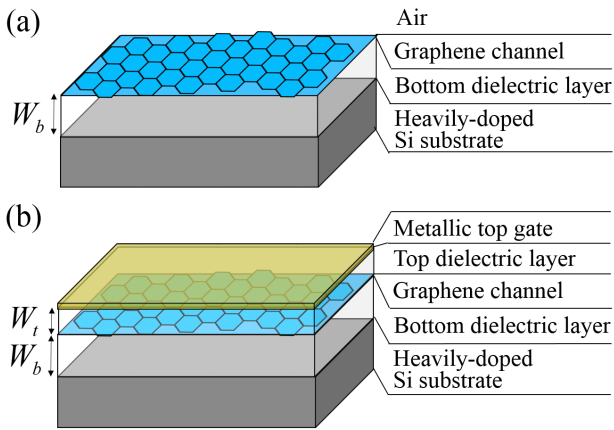}
  \caption{\label{FigSchematics}
    Schematic views of (a) an ungated graphene structure with a 
    heavily-doped Si substrate where the top surface is exposed on
    the air and (b) a gated graphene structure with a
    heavily-doped Si substrate and a metallic top gate.}
  \end{center}
\end{figure}

We investigate plasmons in
an ungated graphene structure with a heavily-doped p$^{+}$-Si 
substrate, where the graphene layer is exposed on the air,
as well as
a gated graphene structure with the substrate
and a metallic top gate, which are schematically
shown in Figs.~\ref{FigSchematics}(a) and (b), respectively. 
The thickness of
the substrate is assumed to be sufficiently larger than the skin
depth of the substrate surface plasmons.
The top gate can be considered as perfectly conducting metal,
whereas the heavily-doped Si substrate is characterized by
its complex dielectric constant.

Here, we use the hydrodynamic equations to describe the electron
motion in graphene~\cite{Svintsov2012}, while using the simple
Drude model for the hole motion in the substrate
(due to virtual independence of the effective mass in the substrate
on the electron density, in contrast to graphene).
In addition, these are accompanied by the self-consistent
2D Poisson equation
(The formulation used here almost follows that for compound 
semiconductor high-electron-mobility transistors, 
see Ref.~\onlinecite{Satou2004b}).
Differences are the hydrodynamic equations
accounting for the linear dispersion of graphene
and material parameters of the substrate and 
dielectric layers.
In general, the existence of both electrons and holes in graphene
results in various modes such as electrically passive electron-hole 
sound waves in intrinsic graphene as well as in huge damping of 
electrically active modes due to the electron-hole friction, 
as discussed in Ref.~\onlinecite{Svintsov2012}.
Here, we focus on
the case where the electron concentration is much higher than
the hole concentration and therefore the damping associated with 
the friction can be negligibly small.
Besides, for the generalization purpose, we formulate the
plasmon dispersion equation for the gated structure; that for
the ungated structure can be readily found by taking the
limit $W_{t}\to\infty$ (see Fig.~\ref{FigSchematics}).

Then, assuming the solutions of the form
$\exp(ikx-i\omega t)$, where $k=2\pi/\lambda$ and $\omega$ are 
the plasmon wavenumber and frequency ($\lambda$ denotes the 
wavelength), the 2D Poisson equation coupled with the linearized
hydrodynamic equations can be expressed as follows:
\begin{equation}\label{EqPoissonHydrodynamic}
  \frac{\p^{2}\varphi_{\omega}}{\p z^{2}}
  -k^{2}\varphi_{\omega}
  = -\frac{8\pi e^{2}\Sigma_{e}}{3m_{e}\epsilon}
  \frac{k^{2}}{\omega^{2}+i\nu_{e}\omega-\frac{1}{2}(\vF k)^{2}}
  \varphi_{\omega}\delta(z),
\end{equation}
where $\varphi_{\omega}$ is the ac (signal) component of
the potential,
$\Sigma_{e}$, $m_{e}$, and $\nu_{e}$ are the steady-state electron 
concentration, the hydrodynamic ``fictitious mass'', 
and the collision frequency in graphene, respectively,
and $\epsilon$ is the dielectric constant which is different in
different layers.
The electron concentration and fictitious mass are related to
each other through the electron Fermi level, $\mu_{e}$, 
and electron temperature, $T_{e}$:
\begin{equation}\label{EqConcentration}
  \Sigma_{e} = \int_0^\infty \frac{2\varepsilon}{\pi\hbar^{2}\vF^{2}}
  \left[1+\exp\left(\frac{\varepsilon-\mu_e}{\kB T_{e}}\right)
    \right]^{-1} d\varepsilon,
\end{equation}
\begin{equation}\label{EqFictitiousMass}
  m_{e} = \frac{1}{\vF^{2} \Sigma_{e}} 
  \int_0^\infty\frac{2\varepsilon^{2}}{\pi\hbar^{2}\vF^{2}}
  \left[1+\exp(\frac{\varepsilon-\mu_e}{\kB T_{e}})\right]^{-1}.
\end{equation}
In the following we fix $T_{e}$ and treat the fictitious mass
as a function of $\Sigma_{e}$.
The dielectric constant can be represented as
\begin{equation}\label{EqDielectricConstant}
  \epsilon = \left\{
  \begin{array}{ll}
    \epst, & 0<z<W_{t}, \\
    \epsb, & -W_{b}<z<0, \\
    \epss[1-\Omega_{s}^{2}/\omega(\omega+i\nu_{s})], & z<-W_{b}, \\
  \end{array}
  \right.
\end{equation}
where $\epst$, $\epsb$, and $\epss$ are the {\it static} dielectric 
constants of the top and bottom dielectric layers and the substrate,
respectively,
$\Omega_{s} = \sqrt{4\pi e^{2}N_{s}/m_{h}\epss}$ is
the bulk plasma frequency in the substrate with
$N_{s}$ and $m_{h}$ being the doping concentration and hole 
effective mass,
and $\nu_{s}$ is the collision frequency in the substrate,
which depends on the doping concentration.
The dielectric constant in the substrate is a sum of the
static dielectric constant of Si, $\epss=11.7$ 
and the contribution from the
Drude conductivity. The dependence of the collision frequency,
$\nu_{s}$, on the doping concentration, $N_{s}$, is calculated
from the experimental data for the hole mobility
at room temperature in Ref.~\onlinecite{Bulucea-SSE-1993}.

We use the following boundary conditions:
 vanishing potential at the gate and
far below the substrate, $\varphi_{\omega}|_{z=W_{t}} = 0$ and 
$\varphi_{\omega}|_{z=-\infty} = 0$; continuity conditions
of the potential at interfaces between different layers,
$\varphi_{\omega}|_{z=+0} = \varphi_{\omega}|_{z=-0}$
and $\varphi_{\omega}|_{z=-W_{b}+0} = \varphi_{\omega}|_{z=-W_{b}-0}$;
a continuity condition of the electric flux density 
at the interface between the bottom dielectric layer
and the substrate in the $z$-direction,
$\epsb\p\varphi_{\omega}/\p z|_{z=-W_{b}+0} =
\epss\p\varphi_{\omega}/\p z|_{z=-W_{b}-0}$; and a
jump of the electric flux density at the graphene layer,
which can be derived from Eq.~(\ref{EqPoissonHydrodynamic}).
Equation~(\ref{EqPoissonHydrodynamic}) together with
these boundary conditions yield the following dispersion equation
\begin{equation}\label{EqDisp}
  F_{gr}(\omega)F_{sub}(\omega) = A_{c},
\end{equation}
where
\begin{eqnarray}
  & & F_{gr}(\omega) = \omega^{2}+i\nu_{e}\omega
  -\frac{1}{2}(\vF k)^{2}-\Omega_{gr}^{2},
  \label{EqFgr} \\
  & & F_{sub}(\omega) = \omega(\omega+i\nu_{s})-\Omega_{sub}^{2},
  \label{EqFsub} \\
  & & A_{c} = \frac{\epsb^{2}(\Hb^{2}-1)}
        {(\epsb\Hb+\epst\Ht)(\epss+\epsb\Hb)}
        \Omega_{gr}^{2}\Omega_{sub}^{2},
        \label{EqAc}
\end{eqnarray}
\begin{equation}\label{EqFreqGr}
  \Omega_{gr} =
  \sqrt{\frac{8\pi e^{2}\Sigma_{e}k}{3m_{e}\epsilon_{gr}(k)}}, \
  \epsilon_{gr}(k) = 
  \epst\Ht+\epsb\frac{\epsb+\epss\Hb}{\epss+\epsb\Hb},
\end{equation}
\begin{equation}\label{EqFreqSub}
  \Omega_{sub} = 
  \sqrt{\frac{4\pi e^{2}N_{s}}{m_{h}\epsilon_{sub}(k)}}, \
  \epsilon_{sub}(k) = 
  \epss+\epsb\frac{\epsb+\epst\Ht\Hb}{\epsb\Hb+\epst\Ht},
\end{equation}
and $H_{b,t} = \coth kW_{b,t}$.
In Eq.~(\ref{EqDisp}), the term $A_{c}$ on the right-hand side
represents the coupling between graphene plasmons and substrate
surface plasmons. If $A_{c}$ were zero, the equations 
$F_{gr}(\omega)=0$ and $F_{sub}(\omega)=0$ would give independent
dispersion relations for the former and latter, respectively.
Qualitatively, Eq.~(\ref{EqAc}) indicates that the coupling 
occurs unless $kW_{b}\gg1$ or $kW_{t}\ll1$, i.e., unless
the separation of the graphene channel and the substrate is
sufficiently large or the gate screening of graphene plasmons is
effective.
Note that the non-constant frequency dispersion of the
substrate surface plasmon in Eq.~(\ref{EqFreqSub})
is due to the gate screening, which is similar to that in
the structure with two parallel metal electrodes~\cite{Maier}.
Equation~(\ref{EqDisp}) yields two modes
which have dominant potential distributions near the graphene channel 
and inside the substrate, respectively.
Hereafter, we focus on the
oscillating mode primarily in the graphene channel; we call it
``channel mode'', whereas we call the other mode ``substrate mode''.

\section{Ungated Plasmons}

First, we study plasmons in the ungated structure.
Here, the temperature, electron concentration, and 
collision frequency in graphene are fixed to
$T_{e} = 300$ K, $\Sigma_{e} = 10^{12}$ cm$^{-2}$,
 and $\nu_{e} = 3\times10^{11}$ s$^{-1}$.
With these values of the temperature and
concentration the fictitious mass is equal to $0.0427 m_{0}$, where
$m_{0}$ is the electron rest mass. The value of the
collision frequency is typical to the acoustic-phonon scattering
at room temperature~\cite{Hwang2008b}.
As for the structural parameters,
we set $\epst = 1$ and $W_{t}\to\infty$, and
we assume an SiO$_{2}$ bottom dielectric layer with
$\epsb = 4.5$. Then Eq.~(\ref{EqDisp}) is solved numerically.

\begin{figure}[t]
  \begin{center}
    \includegraphics[width=7.5cm]
                    {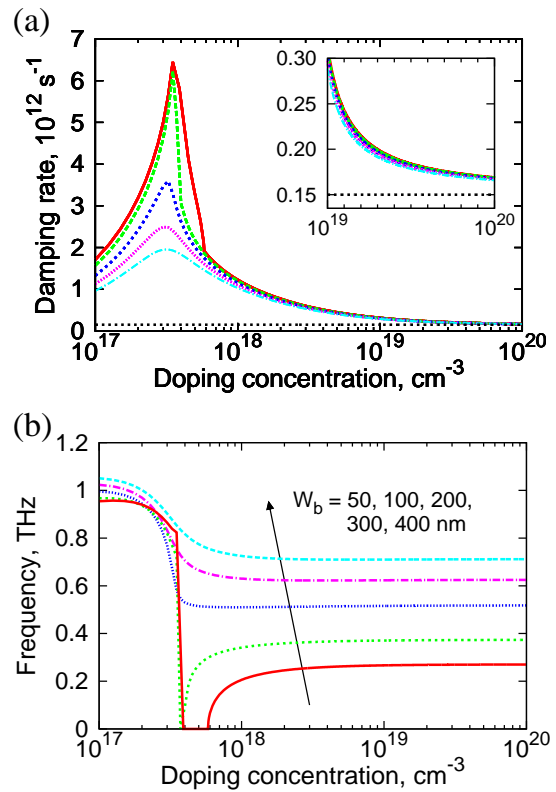}
  \caption{\label{FigFreqDampUngated-concDep}
  Dependences of (a) the plasmon damping rate and (b) frequency
  on the substrate doping concentration, $N_{s}$, with the plasmon
  wavenumber $k = 14\times10^{3}$ cm$^{-1}$ (the wavelength 
  $\lambda = 4.5$ $\mu$m) and with different thicknesses
  of the bottom dielectric layer, $W_{b}$, in the ungated graphene 
  structure.
  The inset in (a) shows the damping rate in the range
  $N_{s}=10^{19}-10^{20}$ cm$^{-3}$ (in linear scale).}
  \end{center}
\end{figure}

Figures~\ref{FigFreqDampUngated-concDep}(a) and (b) 
show the dependences of the plasmon 
damping rate and frequency on the substrate doping concentration
with the plasmon wavenumber $k = 14\times10^{3}$ cm$^{-1}$
(i.e., the wavelength $\lambda = 4.5$ $\mu$m) and with 
different thicknesses of the bottom dielectric layer, $W_{b}$.
The value of the plasmon wavelength is chosen so that 
it gives the frequency around $1$ THz in the limit $N_{s}\to0$.
They clearly demonstrate that there is a huge resonant increase in the
damping rate at around $N_{s} = 3\times10^{17}$ cm$^{-3}$
as well as a drop of the frequency. This is the manifestation of 
the resonant coupling of the graphene plasmon
and the substrate surface plasmon.
The resonance corresponds
to the situation where the frequencies of graphene plasmons
and substrate surface plasmons coincide, in other words,
where the exponentially decaying tail of electric field of graphene 
plasmons resonantly excite the substrate surface plasmons. 

%
%
%

At the resonance, the damping rate becomes larger than $10^{12}$
s$^{-1}$, over $10$ times larger than the contribution
from the acoustic-phonon scattering in graphene, 
$\nu_{e}/2 = 1.5\times10^{11}$ s$^{-1}$.
For structures with $W_{b} = 50$ and $100$ nm, even the damping rate
is so large that the frequency is dropped down to zero; this
corresponds to an overdamped mode.
It is seen in Figs.~\ref{FigFreqDampUngated-concDep}(a) and
(b) that the coupling effect becomes weak as the 
thickness of the bottom dielectric layer increases.
The coupling strength at the resonance is determined by the ratio
of the electric fields at the graphene layer and at the interface 
between the bottom dielectric layer and substrate.
In the case of the ungated structure with a relatively low
doping concentration,
it is roughly equal to $\exp(-kW_{b})$. 
Since $\lambda = 4.5$ $\mu$m is much larger
than the thicknesses of the bottom dielectric layer in the structures 
under consideration, i.e.,
$kW_{b} \ll 1$, the damping rate and frequency 
in Figs.~\ref{FigFreqDampUngated-concDep}(a) and (b)
exhibit the rather slow dependences on the thickness.

Away from the resonance, we have several nontrivial features
in the concentration dependence of the damping rate.
On the lower side of the doping concentration, the damping
rate increase does not vanish until $N_{s} = 10^{14}-10^{15}$
cm$^{-3}$.
This comes from the wider field spread of the channel mode
into the substrate due to the ineffective screening by the
low-concentration holes.
On the higher side, one can also see 
a rather broad linewidth of the resonance with respect to the doping 
concentration, owning to the large, concentration-dependent damping 
rate of
the substrate surface plasmons, and a contribution
to the damping rate is not negligible even when the doping 
concentration is increased two-orders-of-magnitude higher.
In fact, with $N_{s} = 10^{19}$ cm$^{-3}$, the damping rate is 
still twice larger than the contribution
from the acoustic-phonon scattering. 
The inset in Fig.~\ref{FigFreqDampUngated-concDep}(a) indicates that
the doping concentration must be at least larger than
$N_{s} = 10^{20}$ cm$^{-3}$ for the coupling effect to be
smaller than the contribution from the 
acoustic-phonon scattering, although the latter is still 
nonnegligible.
It is also seen from the inset that, with very high doping 
concentration,
the damping rate is almost insensitive to $W_{b}$.
This originates from the screening by the substrate that
strongly expands the field spread into the bottom dielectric layer.

As for the dependence of the frequency, it tends to 
a lower value in the limit $N_{s}\to\infty$ than that in the limit 
$N_{s}\to0$, as seen
in Figure~\ref{FigFreqDampUngated-concDep}(b), along with
the larger dependence on the thickness $W_{b}$.
This corresponds to the transition of the channel mode from an 
ungated plasmon mode to a gated plasmon mode,
where the substrate effectively acts as a back gate.

\begin{figure}[t]
  \begin{center}
    \includegraphics[width=8cm]{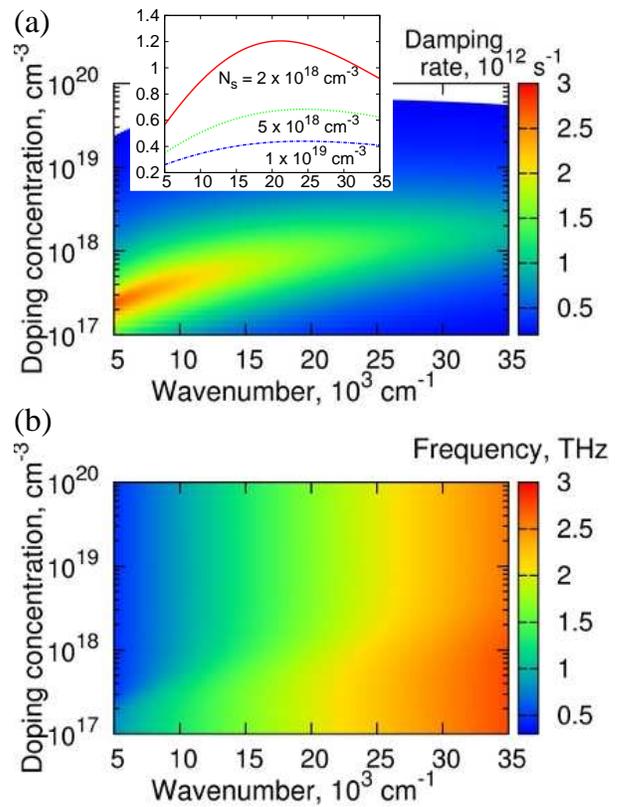}
  \caption{\label{FigFreqDampUngated-2DDep}
  Dependences of (a) the plasmon damping rate and (b) frequency
  on the substrate doping concentration, $N_{s}$, and the plasmon
  wavenumber, $k$, with different the thickness of the
  bottom dielectric layer
  $W_{b} = 300$ nm in the ungated graphene structure.
  The inset of (a) shows the wavenumber dependence of the damping
  rate with certain doping concentrations.
  The region with the damping rate below $0.2\times10^{12}$ s$^{-1}$
  is filled with white in (a).}
  \end{center}
\end{figure}

To illustrate the coupling effect with various frequencies in the
THz range, dependences of the plasmon damping 
rate and frequency on the substrate doping concentration and plasmon 
wavenumber with $W_{b} = 300$ nm are plotted in 
Figs.~\ref{FigFreqDampUngated-2DDep}(a) and (b).
In Fig.~\ref{FigFreqDampUngated-2DDep}(a),
the peak of the damping rate shifts to the higher doping 
concentration as the wavenumber increases, whereas its value
decreases. The first feature can be understood from the
matching condition of the 
wavenumber-dependent frequency of the ungated graphene plasmons 
and the doping-concentration-dependent frequency of the
substrate surface plasmons, i.e., 
$\Omega_{gr}\propto k^{1/2}$, roughlly speaking, and 
$\Omega_{sub}\propto N_{s}^{1/2}$. The second feature 
originates from the exponential decay factor, $\exp(-kW_{b})$, 
of the electric field of the channel mode at the interface
between the bottom dielectric layer and the substrate;
since the doping concentration is $\lesssim 10^{18}$ cm$^{-3}$
at the resonance for any wavevector in 
Fig.~\ref{FigFreqDampUngated-2DDep}, the exponential decay is 
valid.
Also, with a fixed doping concentration, say
$N_{a} > 10^{19}$ cm$^{-3}$, the damping rate has a maximum at a 
certain wavenumber,
resulting from the first feature
(see the inset in Fig.~\ref{FigFreqDampUngated-2DDep}(a)).


\section{Gated Plasmons}

Next, we study plasmons in the gated structures.
We consider the same electron concentration, ficticious mass, and 
collision frequency, $\Sigma_{e} = 10^{12}$ cm$^{-2}$, 
$m_{e} = 0.0427 m_{0}$, and $\nu_{e} = 3\times10^{11}$
s$^{-1}$, as the previous section.
As examples of materials for top/bottom dielectric layers,
we examine Al$_{2}$O$_{3}$/SiO$_{2}$ and 
diamond-like carbon (DLC)/3C-SiC. These materials choices
not only reflect the realistic combination of dielectric materials
available today, but also demonstrate two distinct situations
for the coupling effect under consideration, where
$\epst > \epsb$ for the former and $\epst < \epsb$ for the latter.

\begin{figure}[t]
  \begin{center}
    \includegraphics[width=8.75cm]
                    {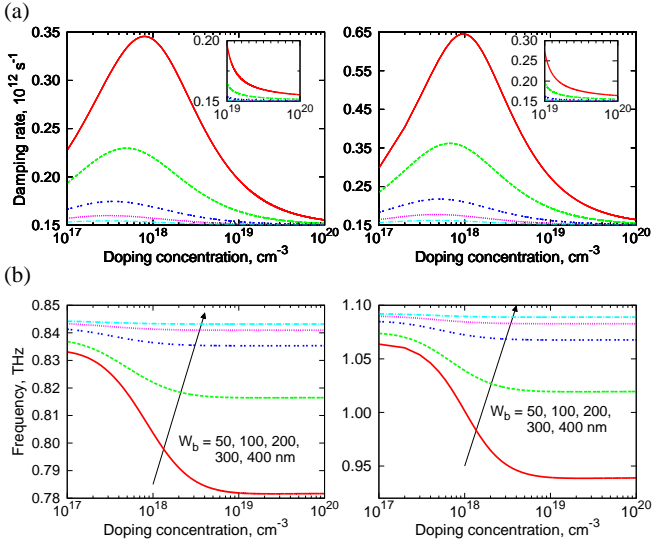}
  \caption{\label{FigFreqDampGated-concDep-SiO2}
  Dependences of (a) the plasmon damping rate and (b) frequency
  on the substrate doping concentration, $N_{s}$, with the plasmon
  wavelength $\lambda = 1.7$ $\mu$m (the wavenumber $k=37\times10^{3}$ 
  cm$^{-1}$), with thicknesses of the Al$_{2}$O$_{3}$ top 
  dielectric layer $W_{t} = 20$ and $40$ nm (left and right panels,
  respectively), and with different thicknesses of the SiO$_{2}$ 
  bottom dielectric layer, $W_{b}$, in the gated graphene structure.
  The insets in (a) show the damping rate in the range
  $N_{s}=10^{19}-10^{20}$ cm$^{-3}$ (in linear scale).}
  \end{center}
\end{figure}
\begin{figure}[t]
  \begin{center}
    \includegraphics[width=8cm]{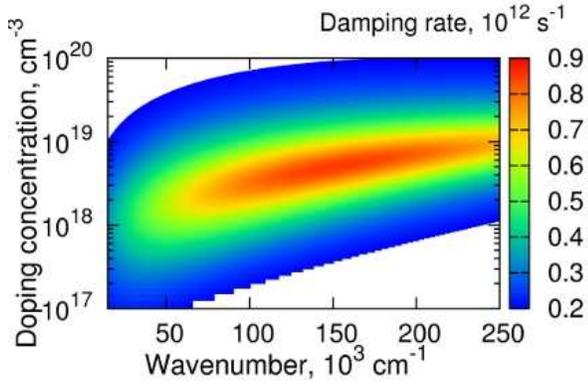}
    \caption{\label{FigDampGated-2DDep-SiO2}
      Dependence of the plasmon damping rate on the substrate doping 
      concentration, $N_{s}$, and the plasmon wavenumber, $k$, 
      with different the thicknesses of the Al$_{2}$O$_{3}$ top 
      dielectric layer $W_{t} = 20$ nm and 
      the SiO$_{2}$ bottom dielectric 
      layer $W_{b} = 50$ nm in the gated graphene structure.
      The region with the damping rate below
      $0.2\times10^{12}$ s$^{-1}$ is filled with white.}
  \end{center}
\end{figure}

Figures~\ref{FigFreqDampGated-concDep-SiO2}(a) and (b) 
show the dependences of the plasmon 
damping rate and frequency on the substrate doping concentration
with the wavenumber $k=37\times10^{3}$ cm$^{-1}$
(the plasmon wavelength $\lambda = 1.7$ $\mu$m), with
thicknesses of the Al$_{2}$O$_{3}$ top dielectric layer 
$W_{t} = 20$ and $40$ nm, and with different thicknesses of the 
SiO$_{2}$ bottom dielectric layer, $W_{b}$.
As seen, the resonant peaks in the damping rate as well as the 
frequency drop due to the coupling effect appear, 
although the peak values are substantially smaller
than those in the ungated structure 
(cf. Fig.~\ref{FigFreqDampUngated-concDep}).
The peak value decreases rapidly as the thickness of the bottom
dielectric layer increases; it almost vanishes
when $W_{b} \geq 300$ nm.
These reflect the fact that in the gated structure the electric 
field of the channel mode is confined dominantly in the top 
dielectric layer due to the gate screening effect.
The field only 
weakly spreads into the bottom dielectric layer,
where its characteristic length is roughtly proportional to 
$W_{t}$, rather than the wavelength $\lambda$ as in the 
ungated structure.
Thus, the coupling effect on the damping rate together with
on the frequency vanishes quickly as $W_{b}$ increases, even when
the wavenumber is small and $kW_{b} \ll 1$.
More quantitatively, the effect is negligible when 
the first factor of $A_{c}$ given in Eq.~(\ref{EqAc})
in the limit $k W_{b} \ll 1$ and $k W_{t} \ll 1$,
\begin{equation}\label{EqFirstFactor}
  \frac{\epsb^{2}(\Hb^{2}-1)}{(\epsb\Hb+\epst\Ht)(\epss+\epsb\Hb)}
  \simeq \frac{1}{1+(W_{b}/\epsb)/(W_{t}/\epst)}
\end{equation}
is small, i.e., when the factor 
$(W_{b}/\epsb)/(W_{t}/\epst)$ is much larger than unity.
A rather strong dependence of the damping rate on $W_{b}$
can be also seen with high doping concentration, in the
insets of Fig.~(\ref{FigFreqDampGated-concDep-SiO2})(a).

Figure~\ref{FigDampGated-2DDep-SiO2} shows
the dependence of the plasmon damping rate on the 
substrate doping concentration and plasmon wavenumber, with
dielectric layer thicknesses
$W_{t} = 20$ and $W_{b} = 50$ nm.
As compared with the case of the ungated structure 
(Fig.~\ref{FigFreqDampUngated-2DDep}(a)), 
the peak of the damping rate exhibits a different
wavenumber dependence; it shows a
broad maximum at a certain wavenumber (around $150\times10^{3}$ 
cm$^{-1}$ in
Fig.~\ref{FigDampGated-2DDep-SiO2})
unlike the case of the ungated structure, where
the resonant peak decreases monotonically as increasing 
the wavenumber. This can be explained by the screening effect of the
substrate against that of the top gate.
When the wavenumber is small and the doping concentration 
corresponding to the resonance is low, the field created
by the channel mode is mainly screened by the gate and the field
is weakly spread into the bottom direction.
As the doping concentration increases (with increase in the
wavevector which gives the resonance),
the substrate begins to act as a back gate and
the field spreads more into the bottom dielectric layer,
so that the coupling effect becomes stronger.
When the wavenumber becomes so large that $k W_{b} \ll 1$
does not hold, the field spread is no longer governed dominantly by
the substrate or gate screening, i.e., the channel mode begins to be 
``ungated'' by the substrate. Eventually, the coupling effect 
on the damping rate again becomes weak,
with the decay of the field being proportional to
$\exp(-k W_{b})$.

\begin{figure}[t]
  \begin{center}
    \includegraphics[width=8.75cm]
                    {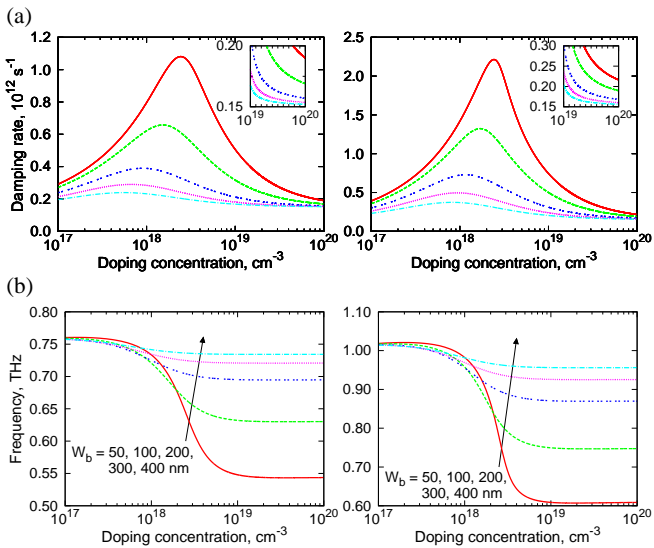}
  \caption{\label{FigFreqDampGated-concDep-GOS}
    The same as Figs.~\ref{FigFreqDampGated-concDep-SiO2}(a) and
    (b) but with the DLC top  and 3C-SiC bottom dielectric layer.}
  \end{center}
\end{figure}

\begin{figure}[t]
  \begin{center}
    \includegraphics[width=8cm]{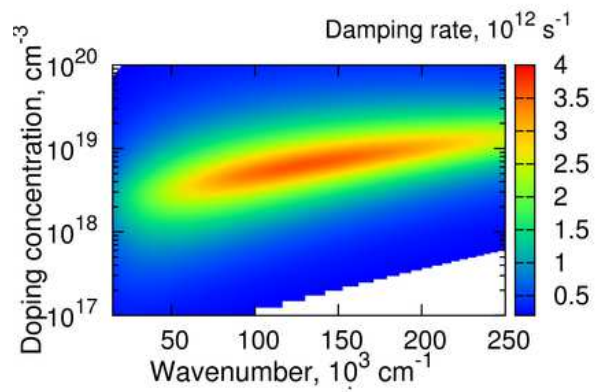}
    \caption{\label{FigDampGated-2DDep-GOS}
    The same as Fig.~\ref{FigDampGated-2DDep-SiO2}
    but with the DLC top and 3C-SiC bottom dielectric layer.}
  \end{center}
\end{figure}

As illustrated in Eq.~(\ref{EqFirstFactor}), the coupling effect in 
the gated strcture is characterized by the factor 
$(W_{b}/\epsb)/(W_{t}/\epst)$ when the conditions
$k W_{b} \ll 1$ and $k W_{t} \ll 1$ are met.
This means that not only the thicknesses of the dielectric layers
but also their dielectric constants are very important parameters 
to determine the coupling strength.
For example,
if we adapt the high-k material, e.g., HfO$_{2}$ in the top
dielectric layer, it results in the more effective gate screening
than in the gated structure with the Al$_{2}$O$_{3}$ top
dielectric layer, so that the coupling effect can be suppressed
even with the same layer thicknesses.
The structure with the DLC top and 3C-SiC bottom dielectric layers
(with $\epst=3.1$~\cite{Takakuwa} and $\epsb=9.7$) 
corresponds to the quite 
opposite situation, where the gate screening becomes weak and
the substrate screening becomes more effective, so that
stronger coupling effect is anticipated.
Figures~\ref{FigFreqDampGated-concDep-GOS}(a), (b), and 
\ref{FigDampGated-2DDep-GOS}
show the same dependences as in 
Figs~\ref{FigFreqDampGated-concDep-SiO2}(a), (b), and 
\ref{FigDampGated-2DDep-SiO2}, respectively,
for the structure with the DLC top 
and 3C-SiC bottom dielectric layers. Comparing with those for
the structure with the Al$_{2}$O$_{3}$ top 
and SiO$_{2}$ bottom dielectric layers, the damping rate as well
as the frequency are more influenced by the coupling effect
in the entire ranges of the doping concentration and wavevector.
In particular, the increase in the damping rate with high doping 
concentration $N_{s} = 10^{19}-10^{20}$ cm$^{-3}$ and the thickness 
of the bottom layer $W_{b} = 50-100$ nm, which are typical values
in real graphene samples, is much larger.
However, this increase can be avoided by adapting thicker
bottom layer, say, $W_{b} \gtrsim 200$ nm or by increasing
the doping concentration.

%

\section{Conclusions}

In summary, we studied theoretically the coupling of plasmons in 
graphene at THz frequencies with surface plasmons in a heavily-doped 
substrate.
We demonstrated that in the ungated graphene structure
there is a huge resonant increase in the damping rate of
the ``channel mode'' at a certain doping concentration of the
substrate ($\sim10^{17}$ cm$^{-2}$) and the increase can be more than 
$10^{12}$ s$^{-1}$,
due to the resonant coupling of the graphene plasmon and the
substrate surface plasmon. The dependences of the damping rate 
on the doping concentration, the thickness of the bottom dielectric 
layer, and the plasmon wavenumber are associated with
the field spread of the channel mode into the bottom dielectric
layer and into the substrate.
We revealed that even with very high doping 
concentration ($10^{19}-10^{20}$ cm$^{-2}$), away from the resonance, 
the coupling effect causes nonnegligible increase in the damping rate
compared with the acoustic-phonon-limited damping rate.
In the gated graphene structure, the coupling effect can be much 
reduced compared with that in the ungated structure, 
reflecting the fact that the field is confined dominantly 
in the top dielectric layer due to the gate screening. However,
with very high doping concentration, it was shown that
the screening by the substrate effectively spreads the field
into the bottom dielectric layer and the increase in the damping rate
can be nonnegligible. These results suggest that the structural 
parameters such as the thicknesses and dielectric constants of the 
top and bottom dielectric layers must be properly chosen for the THz 
plasmonic devices in order to reduce the coupling effect.

\begin{acknowledgments}
Authors thank M. Suemitsu and S. Sanbonsuge for providing 
information about the graphene-on-silicon structure and
Y. Takakuwa, M. Yang, H. Hayashi, and T. Eto for providing
information about the diamond-like-carbon dielectric layer.
This work was supported by JSPS Grant-in-Aid for Young Scientists
(B) (\#23760300),
by JSPS Grant-in-Aid for Specially Promoted Research (\#23000008),
and by JST-CREST.
\end{acknowledgments}



\begin{thebibliography}{99}
\bibitem{Takakuwa}
  The dielectric constant of DLC varies in the range between $3.1$
  and $7.8$, depending on its growth condition~\cite{Hayashi}.
  Here, we choose the lowest value for demonstration of the
  case where $\epst \ll \epsb$.
\end{thebibliography}
\end{document}